\documentclass[aps,twocolumn,showpacs]{revtex4}
\usepackage{graphicx}

\begin{document}

\title{Complex permittivity of a biased superlattice}
\author{A. Hern\'{a}ndez-Cabrera and P. Aceituno}
\email{ajhernan@ull.es}
\affiliation{Dpto. Fisica Basica, Universidad de La Laguna, La Laguna, 38206-Tenerife,
Spain}
\author{F.T. Vasko}
\email{ftvasko@yahoo.com}
\affiliation{Institute of Semiconductor Physics, NAS Ukraine, Pr. Nauki 41, Kiev, 03028,
Ukraine}
\date{\today}

\begin{abstract}
Intersubband response in a superlattice subjected to a homogeneous electric
field (biased superlattice with equipopulated levels) is studied within the
tight-binding approximation, taking into account the interplay between
homogeneous and inhomogeneous mechanisms of broadening. The complex
dielectric permittivity is calculated beyond the Born approximation for a
wide spectral region and a low-frequency enhancement of the response is
found. A detectable gain below the resonance is obtained for the low-doped $%
GaAs$-based biased superlattice in the THz spectral region. Conditions of
the stimulated emission regime for metallic and dielectric waveguide
structures are discussed. The appearence of a localized THz mode due to BSL
placed at the interface vacuum-dielectric is described.
\end{abstract}

\maketitle

\section{Introduction}

The mechanisms of the stimulated emission due to intersubband transitions of
electrons in different tunnel-coupled structures (monopolar laser effect)
have been investigated during the previous decade (see Refs. in \cite{1,2}).
As a result, both mid-IR and THz lasers viability has been demonstrated with
the use of the scheme based on the vertical transport through quantum
cascade structures, which incorporate injector and active regions. Since a
population inversion appears in the active region, the stimulated emission
occurs for the mode propagated along mid-IR or THz waveguide.

In the case of a biased superlattice (BSL), the vertical current through the
Wannier-Stark ladder, which takes place under the condition $2T\ll
\varepsilon _{\scriptscriptstyle B}$ (here $\varepsilon _{\scriptscriptstyle %
B}/\hbar $ is the Bloch frequency and $T$ stands for the tunneling matrix
element between adjacent\ quantum wells (QWs) \cite{3,4}), does not change
level populations. Thus, the consideration based on the golden rule approach
gives zero absorption. In contrast to this, for the wide miniband BSL, with
the bandwidth $2T\gg \varepsilon _{\scriptscriptstyle B}$, a negative
differential conductivity, i.e. gain due to Bloch oscillations, was studied
theoretically starting the 70s \cite{5} (see further results and references
in \cite{6}) and demonstrated in recent experiments \cite{7,8}. At the same
time, a similar behavior of the THz response, including a crossover from
gain to absorption regime with detuning energy shifted through the
resonance, was reported for the BSL with tight-binding electronic states 
\cite{9}. This contradiction with the simple quantum picture, where a zero
response should take place, and the question about THz gain without
inversion beyond the Born approximation are discussed in \cite{10,11}. To
the best of our knowledge, the consideration of the response of a
tight-binding BSL is not performed yet in spite of both real and imaginary
contributions to the dielectric permittivity, $\Delta \epsilon _{\omega }$,
appear to be essential for wide miniband BSLs \cite{6,8}.

\begin{figure}[tbp]
\begin{center}
\includegraphics{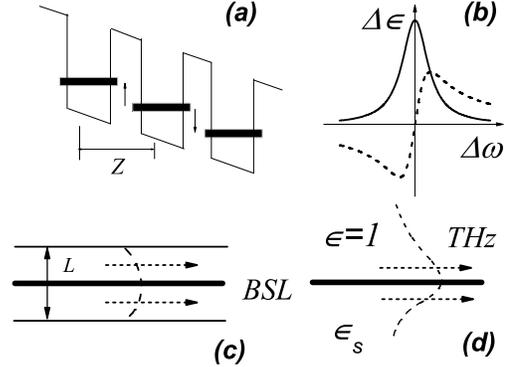}
\end{center}
\par
\addvspace{-1 cm}
\caption{Transitions between broaded levels in BSL of period $Z$ with the
Wannie-Stark ladder $(a)$. Peak and dispersive contributions of these
transitions to the dielectric permittivity, $\Delta \protect\epsilon _{%
\protect\omega }$ (solid and dashed curves are correspondent to $Re(\Delta 
\protect\epsilon _{\protect\omega })$ and $Im(\Delta \protect\epsilon _{%
\protect\omega })$ respectively) $(b)$. Geometries of the THz waveguides
with the BSL placed between the ideal metallic mirrors $(c)$, and at the
interface vacuum-dielectric $(d)$. }
\end{figure}

In this paper, we evaluate the response of a BSL placed on a high-frequency
electric field taking into account both homogeneous and inhomogeneous
mechanisms of broadening exactly. Within the tight-binding approach, which
corresponds to the sequential tunneling picture (Fig. 1$a$), we analyze the
frequency dispersion of complex dielectric permittivity (Fig. 1$b$). The
Green's function formalism is used to describe both homogeneous and
inhomogeneous mechanisms of broadening, and the quasi-equilibrium
distribution of electrons over tunnel-coupled wells. We demonstrate \textit{%
the low-frequency enhancement} of the response under consideration. Further,
the propagation of the transverse magnetic (TM) mode along THz waveguides
with the BSL placed between ideal metallic mirrors (Fig. 1$c$) and at the
interface vacuum-dielectric (Fig. 1$d$) is considered. The appearance of 
\textit{a localized mode} is founded for the second case. We also discuss
the conditions for the stimulated emission regime for the two cases under
consideration.

The paper is organized as follows. In the next section we consider the THz
response of BSL and in Sec. III we discuss the spectral dependencies of
dielectric permittivity, while in Sec. IV we consider the mode propagation
for the BSL placed in the THz waveguide. The last section includes a
discussion of the approximations used and conclusions.

\section{Basic Equations}

Within the framework of the tight-binding approach we describe the
electronic states in BSL using the in-plane inhomogeneous matrix Hamiltonian 
$\hat{h}_{rr^{\prime }}$ and the non-diagonal perturbation matrix due to a
transverse field $[E_{\scriptscriptstyle\bot }\exp (-i\omega t)+c.c.]$
written as $[\widehat{\delta h}_{rr^{\prime }}\exp (-i\omega t)+H.c.]$: 
\begin{eqnarray}
\hat{h}_{rr^{\prime }} &=&\left( \frac{\hat{p}^{2}}{2m}+V_{r\mathbf{x}%
}+r\varepsilon _{\scriptscriptstyle B}\right) \delta _{rr^{\prime
}}+T(\delta _{rr^{\prime }-1}+\delta _{rr^{\prime }+1}),  \nonumber \\
\widehat{\delta h}_{rr^{\prime }} &=&\frac{ev_{\scriptscriptstyle\bot }}{%
\omega }E_{\scriptscriptstyle\bot }(r-r^{\prime })(\delta _{rr^{\prime
}-1}+\delta _{rr^{\prime }+1}).~~~~~~
\end{eqnarray}%
Here $r=0,\pm 1,\ldots $ stands as a QW number, $\hat{p}^{2}/2m$ is the
in-plane kinetic energy operator, and $m$ is the effective mass. The random
potential energy of the $r$-th QW, $V_{r\mathbf{x}}$, is statistically
independent in each QW and includes both short- and long-scale parts of
potential. The Bloch energy, $\varepsilon _{\scriptscriptstyle B}\simeq
|e|FZ $, appears in (1) due to the shift of levels in the SL with period $Z$
under a homogeneous electric field $F$, and $v_{\scriptscriptstyle\bot
}=TZ/\hbar $ stands for the transverse velocity \cite{12}. The
high-frequency current density, $[I_{\omega }\exp (-i\omega t)+c.c.]$, which
involves $\propto \omega ^{-1}$ response and the intersubband contribution
induced by the perturbation $\widehat{\delta h}_{rr^{\prime }}$, is
determined by: 
\begin{equation}
I_{\omega }=i\frac{e^{2}n}{m\omega }E_{\scriptscriptstyle\bot }+i\frac{2ev_{%
\scriptscriptstyle\bot }}{L^{3}}\left\langle \left\langle \sum_{r}\mathrm{sp}%
_{\scriptscriptstyle\Vert }(\widehat{\delta \rho }_{r+1r}-\widehat{\delta
\rho }_{r-1r})\right\rangle \right\rangle ,
\end{equation}%
where $n$ is the electron concentration, factor 2 is due to spin, $\mathrm{sp%
}_{\scriptscriptstyle\Vert }\ldots $ is the trace over in-plane motion, $%
\langle \langle \ldots \rangle \rangle $ is the averaging over random
potentials $V_{r\mathbf{x}}$, and $L^{3}$ is the normalization volume.

The high-frequency contribution to the density matrix in Eq. (2), $[\widehat{%
\delta \rho }_{rr^{\prime }}\exp (-i\omega t)+H.c.]$, is governed by the
independent linearized equations for $\widehat{\delta \rho }_{r}^{%
\scriptscriptstyle(\pm )}\equiv \widehat{\delta \rho }_{r\pm 1r}$, see \cite%
{11, 13}: 
\begin{eqnarray}
&&-i\omega \widehat{\delta \rho }_{r}^{\scriptscriptstyle(\pm )}+\frac{i}{%
\hbar }(\hat{h}_{r\pm 1}\widehat{\delta \rho }_{r}^{\scriptscriptstyle(\pm
)}-\widehat{\delta \rho }_{r}^{\scriptscriptstyle(\pm )}\hat{h}_{r}) 
\nonumber \\
&\simeq &\pm i\frac{ev_{\scriptscriptstyle\bot }}{\hbar \omega }E_{%
\scriptscriptstyle\bot }(\hat{\rho}_{r\pm 1}-\hat{\rho}_{r})
\end{eqnarray}%
with the in-plane Hamiltonian of the $r$-th QW written in the form $\hat{h}%
_{r}=\hat{p}^{2}/2m+V_{r\mathbf{x}}+r\varepsilon _{\scriptscriptstyle B}$.
Here we restrict ourselves to the consideration of only $\propto T^{2}$
contributions and the steady-state density matrix is written as $(\hat{\rho}%
_{o})_{rr^{\prime }}\simeq \delta _{rr^{\prime }}\hat{\rho}_{r}$. Next, we
describe the electron states in the $r$-th QW by the use of the eigenstate
problem $(\hat{p}^{2}/2m+V_{r\mathbf{x}})\psi _{r\mathbf{x}}^{\nu
}=\varepsilon _{r\nu }\psi _{r\mathbf{x}}^{\nu }$, where the quantum number $%
\nu $ marks an in-plane state. Using this basis, we transform Eq. (3) into
the form: 
\begin{eqnarray}
(\varepsilon _{r\pm 1\nu }-\varepsilon _{r\nu ^{\prime }}\pm \varepsilon _{%
\scriptscriptstyle B}-\hbar \omega -i\lambda )\delta \rho _{r}^{%
\scriptscriptstyle(\pm )}(\nu ,\nu ^{\prime })  \nonumber \\
=\pm \frac{ev_{\scriptscriptstyle\bot }}{\omega }E_{\scriptscriptstyle\bot
}[f_{\varepsilon _{r\pm 1\nu }}-f_{\varepsilon _{r\nu ^{\prime }}}]\int d%
\mathbf{x}\psi _{r\pm 1\mathbf{x}}^{\nu }\psi _{r\mathbf{x}}^{\nu ^{\prime
}~\ast }.
\end{eqnarray}%
Here $\lambda \rightarrow +0$ and we apply the quasi-equilibrium
distribution of $r-$th QW, $\hat{\rho}_{r}=f_{\hat{p}^{2}/2m+V_{r\mathbf{x}%
}} $, where $f_{\varepsilon }$ is the Fermi function with chemical
potentials $\mu $, and temperatures $T_{e}$, which are identical for any QW.

The current density (2) is given by 
\begin{eqnarray}
I_{\omega }\simeq i\frac{e^{2}n}{m\omega }E_{\scriptscriptstyle\bot }+i\frac{%
2ev_{\scriptscriptstyle\bot }}{L^{3}}\left\langle \left\langle \sum_{r\nu
\nu ^{\prime }}\left[ \delta \rho _{r}^{\scriptscriptstyle(+)}(\nu ,\nu
^{\prime })\right. \right. \right. ~~~~~~ \\
\left. \left. \left. \times \int d\mathbf{x}\psi _{r+1\mathbf{x}}^{\nu ~\ast
}\psi _{r\mathbf{x}}^{\nu ^{\prime }}-\delta \rho _{r}^{\scriptscriptstyle%
(-)}(\nu ,\nu ^{\prime })\int d\mathbf{x}\psi _{r-1\mathbf{x}}^{\nu ~\ast
}\psi _{r\mathbf{x}}^{\nu ^{\prime }}\right] \right\rangle \right\rangle 
\nonumber
\end{eqnarray}%
and the transverse conductivity, $\sigma _{\omega }^{\scriptscriptstyle\bot
} $, introduced according to the standard formula $I_{\omega }=\sigma
_{\omega }^{\scriptscriptstyle\bot }E_{\scriptscriptstyle\bot }$, takes the
form: 
\begin{eqnarray}
\sigma _{\omega }^{\scriptscriptstyle\bot } &=&i\frac{e^{2}n}{m\omega }+i%
\frac{2(ev_{\scriptscriptstyle\bot })^{2}}{\omega L^{3}}\left\langle
\left\langle \sum_{r\nu \nu ^{\prime }}(f_{\varepsilon _{r+1\nu
}}-f_{\varepsilon _{r\nu ^{\prime }}})\right. \right. \\
&&\times Q_{r+1,r}^{\nu \nu ^{\prime }}\left[ \frac{1}{\varepsilon _{r+1\nu
}-\varepsilon _{r\nu ^{\prime }}+\varepsilon _{\scriptscriptstyle B}-\hbar
\omega -i\lambda }\right.  \nonumber \\
&&\left. \left. \left. +\frac{1}{\varepsilon _{r+1\nu }-\varepsilon _{r\nu
^{\prime }}+\varepsilon _{\scriptscriptstyle B}+\hbar \omega +i\lambda }%
\right] \right\rangle \right\rangle ,  \nonumber
\end{eqnarray}%
where $Q_{r,r^{\prime }}^{\nu \nu ^{\prime }}=\left\vert \int d\mathbf{x}%
\psi _{r\mathbf{x}}^{\nu ~\ast }\psi _{r^{\prime }\mathbf{x}}^{\nu ^{\prime
}}\right\vert ^{2}$ is the overlap factor. We have replaced $r\rightarrow
r+1,\nu \leftrightarrow \nu ^{\prime }$ in the second addendum. In order to
check the non-singular behavior of $\sigma _{\omega }$ at $\omega
\rightarrow 0$, one has to utilize the relation 
\begin{equation}
\frac{n}{m}+\frac{2v_{\scriptscriptstyle\bot }^{2}}{L^{3}}\sum_{r\nu \nu
^{\prime }}Q_{r+1,r}^{\nu \nu ^{\prime }}\frac{f_{\varepsilon _{r+1\nu
}}-f_{\varepsilon _{r\nu ^{\prime }}}}{\varepsilon _{r+1\nu }-\varepsilon
_{r\nu ^{\prime }}+\varepsilon _{\scriptscriptstyle B}}=0,
\end{equation}%
which can be proofed with the use of the definition $n=2\sum_{r\nu
}f_{\varepsilon _{r\nu }}$ and the relation $v_{\scriptscriptstyle\bot
}^{2}Q_{r,r^{\prime}}^{\nu \nu ^{\prime}}=|\langle r\nu |\hat{v}%
_{z}|r^{\prime}\nu ^{\prime}\rangle |^{2}$, see \cite{13A}. Using (7) we
obtain the transverse dielectric permittivity, $\epsilon _{\omega }^{%
\scriptscriptstyle\bot }=\epsilon +i4\pi \sigma _{\omega }^{%
\scriptscriptstyle\bot }/\omega $, in the form: 
\begin{eqnarray}
\epsilon _{\omega }^{\scriptscriptstyle\bot } =\epsilon +\frac{2\pi (2ev_{%
\scriptscriptstyle\bot })^{2}}{\omega ^{2}L^{3}}\left\langle \left\langle
\sum_{r\nu \nu ^{\prime }}Q_{r+1,r}^{\nu \nu ^{\prime }}(f_{\varepsilon
_{r+1\nu }}-f_{\varepsilon _{r\nu ^{\prime }}})\right. \right.  \nonumber \\
\times \left[ \frac{2}{\varepsilon _{r+1\nu }-\varepsilon _{r\nu ^{\prime
}}+\varepsilon _{\scriptscriptstyle B}}-\frac{1}{\varepsilon _{r+1\nu
}-\varepsilon _{r\nu ^{\prime }}+\varepsilon _{\scriptscriptstyle B}-\hbar
\omega -i\lambda }\right.  \nonumber \\
\left. \left. \left. -\frac{1}{\varepsilon _{r+1\nu }-\varepsilon _{r\nu
^{\prime }}+\varepsilon _{\scriptscriptstyle B}+\hbar \omega +i\lambda }%
\right] \right\rangle \right\rangle ~~~~~~
\end{eqnarray}%
and intersubband transitions give a finite contribution to (8) in the static
limit. Finally, we transform \cite{14} the denominators in (8) and rewrite $%
\Delta \epsilon _{\omega }^{\scriptscriptstyle\bot }=\epsilon _{\omega }^{%
\scriptscriptstyle\bot }-\epsilon $ through the spectral density function 
\cite{13,15}, $\mathcal{A}_{r,\varepsilon }(\mathbf{x},\mathbf{x^{\prime }}%
)=\sum_{\nu }\psi _{r\mathbf{x}}^{\nu }\psi _{r\mathbf{x}^{\prime }}^{\nu
~\ast }\delta (\varepsilon -\varepsilon _{r\nu })$, in the following form 
\begin{eqnarray}
\Delta \epsilon _{\omega }^{\scriptscriptstyle\bot } =\frac{2\pi (2ev_{%
\scriptscriptstyle\bot })^{2}}{\omega ^{2}L^{3}}\int_{-\infty }^{\infty
}d\varepsilon \int_{-\infty }^{\infty }d\varepsilon ^{\prime
}(f_{\varepsilon }-f_{\varepsilon ^{\prime }})\left[ \frac{2}{\varepsilon
-\varepsilon ^{\prime }+\varepsilon _{\scriptscriptstyle B}}\right. 
\nonumber \\
\left. -(\varepsilon -\varepsilon ^{\prime }-\Delta \varepsilon
_{-}-i\lambda )^{-1}-(\varepsilon -\varepsilon ^{\prime }+\Delta \varepsilon
_{+}+i\lambda )^{-1}\right] ~~~~~~  \nonumber \\
\times \int d\mathbf{x}\int d\mathbf{x^{\prime }}~\sum_{r}\langle \langle 
\mathcal{A}_{r+1,\varepsilon }(\mathbf{x},\mathbf{x^{\prime }})\mathcal{A}%
_{r,\varepsilon ^{\prime }}(\mathbf{x^{\prime }},\mathbf{x})\rangle \rangle
~~~~
\end{eqnarray}%
with $\Delta \varepsilon _{\pm }=\hbar \omega \pm \varepsilon _{%
\scriptscriptstyle B}$. Thus, we have evaluated the expression for the
transverse response taking into account the scattering processes exactly.

Further, we perform the averaging over short-range and large-scale
potentials in the last factor, keeping in mind that the averaged
characteristics of scattering processes, both for homogeneous and
inhomogeneous mechanisms, do not dependent on the QW number $r$. Following
Eqs. (11 - 13) of Ref. \cite{11} we obtain 
\begin{eqnarray}
&&\int \int \frac{d\mathbf{x}d\mathbf{x^{\prime }}}{L^{3}}\sum_{r}\langle
\langle \mathcal{A}_{r+1,\varepsilon }(\mathbf{x},\mathbf{x^{\prime }})%
\mathcal{A}_{r,\varepsilon ^{\prime }}(\mathbf{x^{\prime }},\mathbf{x}%
)\rangle \rangle  \nonumber \\
&=&\frac{\rho _{\scriptscriptstyle2D}}{2Z}\int_{0}^{\infty }d\xi A(\xi
-\varepsilon )A(\xi -\varepsilon ^{\prime }),
\end{eqnarray}%
where the averaged spectral density, $A(E)$, is written in the integral
form: 
\begin{equation}
A(E)=\int_{-\infty }^{0}\frac{dt}{\pi \hbar }\cos \left( \frac{Et}{\hbar }%
\right) e^{\gamma t/\hbar -(\Gamma t/\hbar )^{2}/2}.
\end{equation}%
Here $\Gamma =\sqrt{\langle w_{r\mathbf{x}}^{2}\rangle }$ is the
inhomogeneous broadening energy due to the large-scale part of the potential
in the $r$-th QW, $w_{r\mathbf{x}}$, and $\gamma $ stands for the
homogeneous broadening energy. In (11) we consider the case of scattering by
zero-radius centers when $\gamma $ does not depend on $\varepsilon $, $p$ or 
$\mathbf{x}$ and the shift of levels, which is logarithmically divergent
without a small-distance cutoff \cite{16}, is included into the zero point
energy. The simple analytical expressions for the spectral density peak take
the form in the limiting cases: 
\begin{equation}
A(E)=\left\{ 
\begin{array}{cc}
\gamma /[\pi (E^{2}+\gamma ^{2})], & \Gamma =0 \\ 
\exp (-E^{2}/2\Gamma ^{2})/(\sqrt{2\pi }\Gamma ), & \gamma =0%
\end{array}%
\right.
\end{equation}%
and $A(E)$ transforms from a Lorentzian towards a Gaussian line shape upon
an increase in the contribution of the inhomogeneous broadening.

According to Eqs. (9, 10), one needs to consider the convolution of spectral
densities, $D(\varepsilon ,\varepsilon ^{\prime })=\int_{0}^{\infty }d\xi
A(\xi -\varepsilon )A(\xi -\varepsilon ^{\prime })$. For the collisionless
case, when (12) is replaced by the $\delta $-function, one should replace $%
D(\varepsilon ,\varepsilon ^{\prime })$ by $\theta (\varepsilon +\varepsilon
^{\prime })\delta (\varepsilon -\varepsilon ^{\prime })$. Modifications of $%
D(\varepsilon ,\varepsilon ^{\prime })$ due to broadening are shown in Fig.
2 for the cases $\gamma =\Gamma $ and $\gamma =3\Gamma $. Using the
variables $\varepsilon -\varepsilon ^{\prime }$ and $(\varepsilon
+\varepsilon ^{\prime })/2$ we obtain $D(\varepsilon ,\varepsilon ^{\prime
}) $ as a peak of width of the order of broadening with respect to $%
\varepsilon -\varepsilon ^{\prime }$ and $D(\varepsilon ,\varepsilon
^{\prime })$ is suppressed at negative $(\varepsilon +\varepsilon ^{\prime
})/2$ \cite{17}. Note that tails of peaks are suppressed if the
inhomogeneous contribution increases.

\begin{figure}[tbp]
\begin{center}
\includegraphics{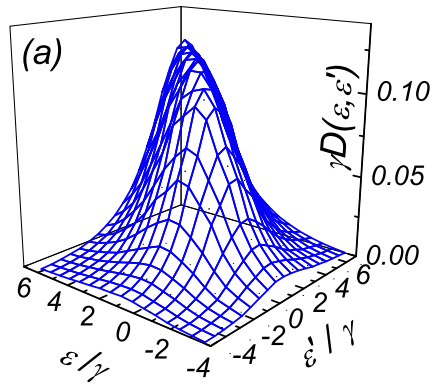} \includegraphics{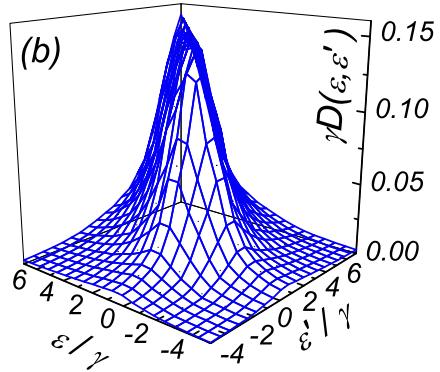}
\end{center}
\par
\caption{(Color online) Dimensionless convolution of spectral densities, $%
\protect\gamma D(\protect\varepsilon ,\protect\varepsilon ^{\prime })$,
plotted for $\protect\gamma =\Gamma $ $(a)$, and $\protect\gamma =3$ $\Gamma 
$ $(b)$. }
\label{Fig. 2}
\end{figure}

\section{Frequency Dispersion}

Here we consider the frequency dispersion of dielectric permittivity tensor $%
\epsilon _{\omega }^{\scriptscriptstyle \| , \bot}$. The in-plane component
of permittivity, $\epsilon _{\omega }^{\scriptscriptstyle \|}$, is written
through the Drude conductivity $\sigma _{\omega}^{\scriptscriptstyle%
\|}=ie^2n/m(\omega + i\gamma /\hbar )$ in the form: 
\begin{equation}
\epsilon _{\omega }^{\scriptscriptstyle \|}=\epsilon -\frac{4\pi e^2n}{%
m\omega (\omega + i\gamma /\hbar )}
\end{equation}
with the homogeneous relaxation frequency, $\gamma /\hbar$, introduced in
Sec. II. The transverse component is given by Eq. (9), which can be written
through $D(\varepsilon ,\varepsilon ^{\prime })$ and the 2D density of
states, $\rho _{\scriptscriptstyle 2D}$, as follows: 
\begin{eqnarray}
\Delta \epsilon _{\omega }^{\scriptscriptstyle\bot} =\frac{\pi (2ev_{%
\scriptscriptstyle\bot })^{2}}{\omega ^{2}Z}\rho _{\scriptscriptstyle%
2D}\int_{-\infty }^{\infty }d\varepsilon \int_{-\infty }^{\infty
}d\varepsilon ^{\prime }~~~~~~~~  \nonumber \\
\times D(\varepsilon ,\varepsilon ^{\prime })(f_{\varepsilon ^{\prime
}}-f_{\varepsilon })\left[ 2/(\varepsilon -\varepsilon ^{\prime
}+\varepsilon _{\scriptscriptstyle B})\right. ~~~~~~~ \\
\left. -(\varepsilon -\varepsilon ^{\prime }-\Delta \varepsilon
_{-}-i\lambda )^{-1}-(\varepsilon -\varepsilon ^{\prime }+\Delta \varepsilon
_{+}+i\lambda )^{-1}\right] .~~~  \nonumber
\end{eqnarray}%
The Fermi distribution, $f_{\varepsilon }$, is connected to the averaged
concentration, $n_{\scriptscriptstyle 2D}=nZ$, according to the standard
relation $n_{\scriptscriptstyle2D}=\int_{-\infty }^{\infty }d\varepsilon
f_{\varepsilon }\rho _{\varepsilon }=\rho _{\scriptscriptstyle%
2D}\int_{-\infty }^{\infty }d\varepsilon f_{\varepsilon }\int_{0}^{\infty
}d\xi A(\xi -\varepsilon )$, where $\rho_{\varepsilon }$ is written through $%
A(E)$ given by Eq. (11).

\begin{figure}[tbp]
\begin{center}
\includegraphics{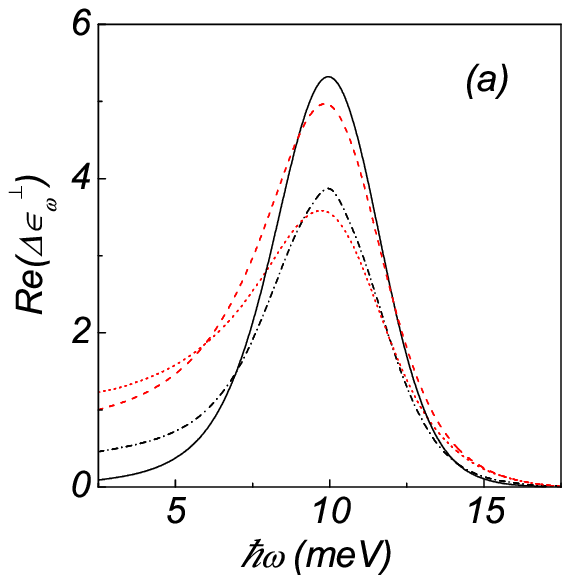} \includegraphics{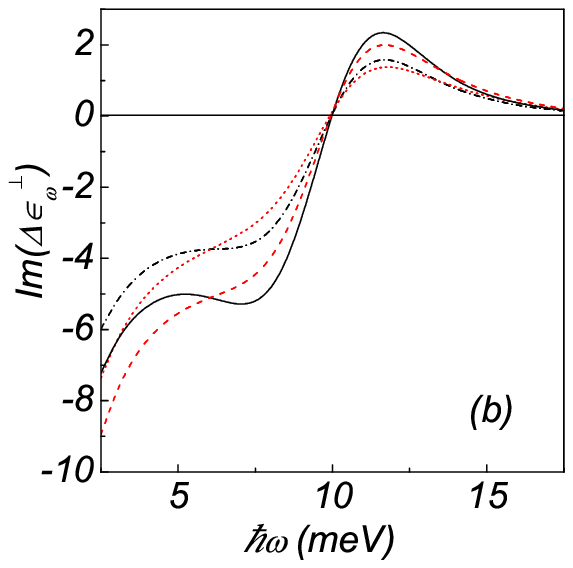}
\end{center}
\par
\addvspace{-1 cm}
\caption{(Color online) Frequency dispersion of the real ($a$) imaginary ($b$%
) parts of dielectric permittivity $\Delta\protect\epsilon _{\protect\omega %
}^{\perp }$ . For the case $\protect\gamma =\Gamma =0.8$ meV one uses $%
\protect\mu =1.5$ meV, $T_{e}=0.5$ meV (solid line) and $\protect\mu =1.5$
meV, $T_{e}=4.5$ meV (dot-dashed line). For the case $\protect\gamma =1.2$
meV, $\Gamma =0.4$ meV one uses $\protect\mu =1.5$ meV, $T_{e}=0.5$ meV
(dashed line) and $\protect\mu =1.5$ meV, $T_{e}=4.5$ meV (dotted line).}
\label{Fig. 3}
\end{figure}

We turn to estimates of the dielectric permittivity for the $%
GaAs/Al_{0.3}Ga_{0.7}As$ BSL with a period $Z=170$ \AA\ and with a tunneling
matrix element $T=2$ meV, which corresponds to a barrier width of $45$ \AA .
The level splitting energy $\varepsilon _{\scriptscriptstyle B}=10$ meV
corresponds to a transverse electric field $F=5.9$ kV/cm. Figure 3 shows the
spectra of $\Delta \epsilon _{\omega }$ for the cases: $\gamma =\Gamma =0.8$
meV, and $\gamma =3\Gamma $ with $\Gamma =0.4$ meV. We have used in
calculations a chemical potential $\mu =1.5$ meV and temperatures $T_{e}=0.5$
meV and $4.5$ meV. The corresponding 2D electron densities are $n_{%
\scriptscriptstyle2D}=6.1\times 10^{10}$ cm$^{-2}$ ($\gamma =\Gamma $ and $%
T_{e}=0.5$ meV), $n_{\scriptscriptstyle2D}=6.8\times 10^{10}$ cm$^{-2}$ ($%
\gamma =3\Gamma $ and $T_{e}=0.5$ meV), and $n_{\scriptscriptstyle%
2D}=1.2\times 10^{11}$ cm$^{-2}$ ($T_{e}=4.5$ meV, both broadening cases).
Both the real (Fig. 3$a$) and imaginary (Fig. 3$b$) parts of $\Delta
\epsilon _{\omega }^{\scriptscriptstyle\bot }$ show a decrease of their peak
value with increasing temperature. A non-symmetric shape of $Re\left( \Delta
\epsilon _{\omega }\right) $ appears due to the singular ($\propto \omega
^{-2}$) factor in Eq.(14). As a result, $Im\left( \Delta \epsilon _{\omega
}\right) $ increases in the low-energy region. Homogeneous broadening, $%
\gamma $, has a bigger influence in the width of the permittivity dispersion
than the inhomogeneous one, $\Gamma $.

\section{Electrodynamics}

Next, using the contribution to dielectric permittivity discussed above, we
consider the in-plane propagation of a THz mode localized at the BSL, as it
is shown in Figs. 1($c,d)$. Since the only $z$-polarized component of the
field is coupled to the BSL, we consider the TM-mode propagating along the $%
OX$-direction with the non-zero components $[E_{\scriptscriptstyle\Vert
}(z)e^{ik_{\omega }x},~0,~E_{\scriptscriptstyle\bot }(z)e^{ik_{\omega }x}]$.
The wave equation for these components can be transformed into the system 
\cite{18}: 
\begin{eqnarray}
\left[ \epsilon _{z\omega }^{\scriptscriptstyle\Vert }\left( \frac{\omega }{c%
}\right) ^{2}+\frac{d^{2}}{dz^{2}}\right] E_{\scriptscriptstyle\Vert }(z)
&=&ik_{\omega }\frac{dE_{\scriptscriptstyle\bot }(z)}{dz}, \\
\left[ \epsilon _{z\omega }^{\scriptscriptstyle\bot }\left( \frac{\omega }{c}%
\right) ^{2}-k_{\omega }^{2}\right] E_{\scriptscriptstyle\bot }(z)
&=&ik_{\omega }\frac{dE_{\scriptscriptstyle\Vert }(z)}{dz},  \nonumber
\end{eqnarray}%
where $\epsilon _{z\omega }^{\scriptscriptstyle\Vert ,\bot }$ is the
dielectric permittivity tensor of the layered media with $\epsilon _{z\omega
}^{\scriptscriptstyle\Vert }$ given by Eq. (13) and $\epsilon _{z\omega }^{%
\scriptscriptstyle\bot }=\epsilon _{z}+\Delta \epsilon _{z\omega }^{%
\scriptscriptstyle\perp }$ given by Eq. (14).

Using the relation $E_{\scriptscriptstyle\bot }(z)=ik_{\omega }[dE_{%
\scriptscriptstyle\Vert }(z)/dz]$ $/[\epsilon _{z}^{\bot }(\omega
/c)^{2}-k_{\omega }^{2}]$ one may write the closed wave equation for $E_{%
\scriptscriptstyle\Vert }(z)$ in the form: 
\begin{equation}
\left[ \frac{d}{dz}\frac{\epsilon _{z\omega }^{\scriptscriptstyle\bot
}(\omega /c)^{2}}{\epsilon _{z}^{\scriptscriptstyle\bot }(\omega
/c)^{2}-k_{\omega }^{2}}\frac{d}{dz}+\epsilon _{z\omega }^{\scriptscriptstyle%
\Vert }\left( \frac{\omega }{c}\right) ^{2}\right] E_{\scriptscriptstyle%
\Vert }(z)=0.
\end{equation}
We consider below three-layer structures with the BSL placed at $|z|<d/2$.
From Eq. (15) one obtains the wave equations with constant coefficients
complemented by boundary conditions at $z\pm d/2$: 
\begin{equation}
\left. \frac{\epsilon _{z}^{\scriptscriptstyle\bot }}{\epsilon _{z}^{%
\scriptscriptstyle\bot }(\omega /c)^{2}-k_{\omega }^{2}}\frac{dE_{%
\scriptscriptstyle\Vert }(z)}{dz}\right\vert _{z=\pm d/2-0}^{z=\pm d/2+0}=0~
\end{equation}%
and the continuity conditions: $\left. E_{\scriptscriptstyle\Vert
}(z)\right\vert _{z=\pm d/2-0}^{z=\pm d/2-0}=0$. In addition, the problem
should be complemented by boundary conditions at $|z|\gg d$. Below we
restrict ourself by the cases of an ideal metallic waveguide and a THz mode
localized at the interface vacuum-dielectric.

\subsection{Metallic Waveguide}

For the ideal metallic waveguide of width $L$ we involve the additional
boundary conditions $E_{\scriptscriptstyle\Vert }(z=\pm L/2)=0$ and the wave
equation (16) takes the form: 
\begin{eqnarray}
\left( \frac{d^{2}}{dz^{2}}+\kappa ^{2}\right) E_{\scriptscriptstyle\Vert
}(z) &=&0,~~|z|>\frac{d}{2}, \\
\left( \frac{d^{2}}{dz^{2}}+\kappa _{\scriptscriptstyle\bot }^{2}\right) E_{%
\scriptscriptstyle\Vert }(z) &=&0,~~|z|<\frac{d}{2},  \nonumber
\end{eqnarray}%
where $\kappa $ and $\kappa _{\bot }$ are determined from $\kappa
^{2}=\epsilon (\omega /c)^{2}-k_{\omega }^{2}$ and $\kappa _{%
\scriptscriptstyle\bot }^{2}=\epsilon ^{\scriptscriptstyle\Vert }(\omega
/c)^{2}-k_{\omega }^{2}\epsilon ^{\scriptscriptstyle\Vert }/\epsilon ^{%
\scriptscriptstyle\bot }$, respectively. Here $\epsilon ^{\scriptscriptstyle%
\Vert ,\bot }$ are the components of the BSL dielectric permittivity and $%
\epsilon $ is the dielectric permittivity of the media inside the waveguide.

We search the asymmetric solution of Eq. (17), which corresponds to the
symmetric transverse field $E_{\scriptscriptstyle\bot }(z)$, in the
following form: 
\begin{equation}
E_{\scriptscriptstyle\Vert }(z)=\left\{ 
\begin{array}{cc}
E_{w}\sin \kappa \left( \frac{L}{2}-z\right) , & \frac{d}{2}<z<\frac{L}{2}
\\ 
E_{\scriptscriptstyle SL}\sin \kappa _{\scriptscriptstyle\bot }z, & |z|<%
\frac{d}{2} \\ 
-E_{w}\sin \kappa \left( \frac{L}{2}+z\right) , & -\frac{d}{2}>z>-\frac{L}{2}%
\end{array}%
\right. ,
\end{equation}%
where the coefficients $E_{w}$ and $E_{\scriptscriptstyle SL}$ are
determined from the above boundary conditions. The solvability condition
gives us the dispersion relation: 
\begin{equation}
\frac{\epsilon }{\kappa }\cos \kappa \frac{L-d}{2}\sin \frac{\kappa _{%
\scriptscriptstyle\bot }d}{2}+\frac{\epsilon ^{\scriptscriptstyle\Vert }}{%
\kappa _{\scriptscriptstyle\bot }}\sin \kappa \frac{L-d}{2}\cos \frac{\kappa
_{\scriptscriptstyle\bot }d}{2}=0,
\end{equation}%
which determines the complex longitudinal wave vector $k_{\omega }$ for the
given BSL parameters and $\omega $.

\begin{figure}[tbp]
\begin{center}
\includegraphics{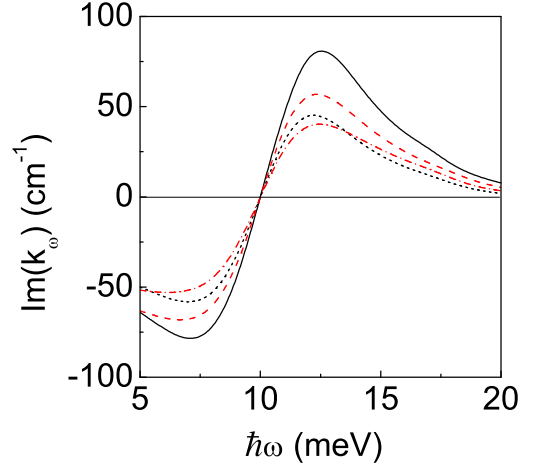}
\end{center}
\par
\addvspace{-1 cm}
\caption{(Color online) Frequency dispersion of $Im(k_{\protect\omega })$
for a metallic waveguide. Curves are marked the same as in Fig.3.}
\label{Fig. 4}
\end{figure}

We solve the dispersion equation (20) using $\epsilon ^{\scriptscriptstyle%
\bot ,\Vert }$ calculated in Sec. III for the BSL width $d=4\mu $m and $%
L=40\mu $m. Considering the lowest mode when the real part of $k_{\omega }$
appears to be close to $\sqrt{\epsilon }\omega /c$, we have plotted $%
Im(k_{\omega })$ in Fig. 4 for the above cases ($\gamma =\Gamma $ and $%
\gamma =3\Gamma $). There are two perfectly defined spectral regions: a gain
region with $Im(k_{\omega })<0$ for $\hbar \omega <\varepsilon _{%
\scriptscriptstyle B}$, and a damping one $Im(k_{\omega })>0$ for $\hbar
\omega >\varepsilon _{\scriptscriptstyle B}$. A marked increase of the gain
can be seen when temperature $T_{e}$ decreases.

\subsection{Localized Mode}

For the case of a BSL placed near the vacuum-dielectric interface, we use
Eqs. (15, 16) with the additional boundary conditions $E_{\scriptscriptstyle%
\Vert }(z\rightarrow \pm \infty )=0$. The wave equations for vacuum, BSL and
substrate regions take the forms: 
\begin{eqnarray}
\left( \frac{d^{2}}{dz^{2}}-\kappa _{v}^{2}\right) E_{\scriptscriptstyle%
\Vert }(z) &=&0,~~z>\frac{d}{2},  \nonumber \\
\left( \frac{d^{2}}{dz^{2}}+\kappa _{\scriptscriptstyle\bot }^{2}\right) E_{%
\scriptscriptstyle\Vert }(z) &=&0,~~|z|<\frac{d}{2}, \\
\left( \frac{d^{2}}{dz^{2}}-\kappa _{s}^{2}\right) E_{\scriptscriptstyle%
\Vert }(z) &=&0,~~z<-\frac{d}{2},  \nonumber
\end{eqnarray}%
where $\kappa _{\scriptscriptstyle\bot }$ is introduced in Sec. IIIA, $%
\kappa _{v}^{2}=(\omega /c)^{2}-k_{\omega }^{2}$ and $\kappa
_{s}^{2}=\epsilon _{s}(\omega /c)^{2}-k_{\omega }^{2}$ with the dielectric
permittivity of substrate, $\epsilon _{s}$. The field distribution is given
by 
\begin{equation}
E_{\scriptscriptstyle\Vert }(z)=\left\{ 
\begin{array}{cc}
E_{+}e^{-\kappa _{v}(z-d/2)}, & z>d/2 \\ 
e_{+}e^{i\kappa _{\scriptscriptstyle\bot }z}+e_{-}e^{-i\kappa _{%
\scriptscriptstyle\bot }z}, & |z|<\frac{d}{2} \\ 
E_{-}e^{\kappa _{s}(z+d/2)}, & z<-d/2%
\end{array}%
\right. ,
\end{equation}%
where $Re(\kappa _{s,v})$ should be positive. The dispersion relation is
obtained from Eqs. (21, 22) in the form: 
\begin{equation}
\frac{\epsilon ^{\scriptscriptstyle\Vert }}{\kappa _{\scriptscriptstyle\bot }%
}\left( \frac{\epsilon _{s}}{\kappa _{s}}+\frac{1}{\kappa _{v}}\right) \cos
\kappa _{\scriptscriptstyle\bot }d-\left( \frac{\epsilon _{s}}{\kappa
_{v}\kappa _{s}}-\frac{(\epsilon ^{\scriptscriptstyle\Vert })^{2}}{\kappa _{%
\scriptscriptstyle\bot }^{2}}\right) \sin \kappa _{\scriptscriptstyle\bot
}d=0.
\end{equation}

\begin{figure}[tbp]
\begin{center}
\includegraphics{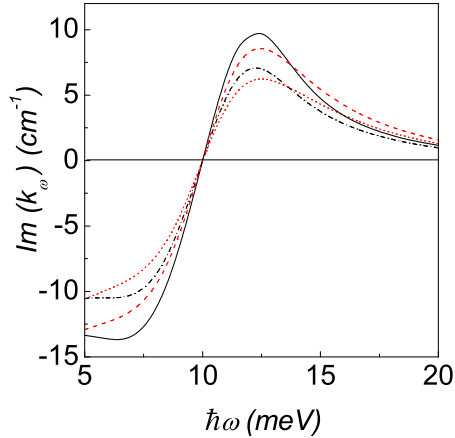}
\end{center}
\par
\addvspace{-1 cm}
\caption{(Color online) Frequency dispersion of $Im(k_{\protect\omega })$
for a dielectric waveguide. Curves are marked the same as in Fig.3.}
\label{Fig. 5}
\end{figure}

The solution of this dispersion equation is performed for the above listed $%
\left( \gamma ,\Gamma \right) $ and $\left( \mu ,T_{e}\right) $ pairs. We
use also the BSL width $d=4\mu $m and $\epsilon _{s}=3.7$, which corresponds
to SiO$_{2}$ substrate. Figure 5 shows $Im(k_{\omega })$ for the cases under
consideration. As in the metallic case, two regions can be selected: a gain
region with $Im(k_{\omega })<0$ for $\hbar\omega <\varepsilon _{%
\scriptscriptstyle B}$, and a damping one where $Im(k_{\omega })>0$ for $%
\hbar \omega >\varepsilon _{\scriptscriptstyle B}$. Contrary to the metallic
case, gain for the dielectric waveguide exists even for low energy values,
at $\hbar\omega <$ 5 meV. Since $Re(\kappa _{v,s})>>Im(\kappa _{v,s})$, the
transverse size of mode is determined by $Re(\kappa _{v,s})^{-1}$ which are
varied from 11.3 $\mu$m at low-frequency region to $Re(\kappa _{v})^{-1}=$%
9.45 $\mu$m and $Re(\kappa _{s})^{-1}=$5.1 $\mu$m at $\hbar \omega =20$ meV.
These results are weakly dependent on temperatures and the broadening cases
under study.

\section{SUMMARY AND CONCLUSIONS}

We have considered here the intersubband response of a biased superlattice
on a THz radiation field beyond the Born approximation. Taking into account
the interplay between homogeneous and inhomogeneous broadening we have
analyzed the spectral and temperature dependencies of the complex dielectric
permittivity in the low-doped BSL. We have found the low-frequency
enhancement of the dispersion of complex dielectric permittivity and have
estimated the conditions to obtain the stimulated emission regime. The
enhancement of the emission due to the THz waveguide effect is also
considered for the cases of the BSL placed between ideal metallic mirrors
and at a vacuum-dielectric interface. The appearence of the localized THz
mode due to BSL placed at the interface vacuum-dielectric is described.

Let us briefly discuss the assumptions used in the present calculations. The
main restriction of the results is the description of the response in the
framework of the tight-binding approach (within an accuracy of the order $%
T^{2}$) which is valid under the condition $\varepsilon _{\scriptscriptstyle %
B}>2T$ and is satisfied for the numerical estimates performed. Note that
beyond the Born approximation the broadening can be comparable to the
averaged electron energy. Next, in spite of the general equations (10) and
(14) are written through an arbitrary spectral density function, with the
use of statistically independent random potentials in each QW, final
calculations were performed for a model that includes scattering by
zero-radius centers and large-scale potential. Such a model describes the
interplay between homogeneous and inhomogeneous mechanisms of broadening .
Other approximations we have made are rather standard. We restrict ourselves
to the case of uniform biased field and QW population neglecting a possible
domain formation caused by the negative differential conductivity of BSL 
\cite{3,19}. One can avoid instabilities in a short enough BSL because the
THz modes propagate in the in-plane directions. The Coulomb correlations,
which modify the response as the concentration increases, are not taken into
account here. This contribution, as well as the consideration of an
intermediate-scale potential, requires a special attention in analogy with
the case of a single QW \cite{20}. Finally, the simplified description of
the ideal (without any damping) waveguide structure is enough to estimate
the characteristic planar size of a device suitable for THz stimulated
emission: one have to compare the maximal negative value of $Im(k_{\omega })$
obtained with a damping length calculated for similar waveguides, see \cite%
{21}. Note, that more complicate waveguides (see Refs. in \cite{22}) may be
effective

To conclude, the simplifications listed do not change the peculiarities of
the THz response or the numerical estimates given in Sec. IV. It seems
likely that the contribution of $Re(\Delta \epsilon _{\omega })$ can be
found experimentally. More detailed numerical simulations are necessary in
order to estimate a potential for applications of BSL as a THz emitter.

\begin{acknowledgments}
This work has been supported in part by Ministerio de Educaci\'{o}n y
Ciencia (Spain) and FEDER under the project FIS2005-01672, and by FRSF of
Ukraine (grant No.16/2).
\end{acknowledgments}


\begin{thebibliography}{99}
\bibitem{1} C. Gmachl, F. Capasso, D.L. Sivco, and A.Y. Cho, Reports on
Progr. in Phys. \textbf{64}, 1533 (2001); J. Faist, D. Hofstetter, M. Beck,
T. Aellen, M. Rochat, and S. Blaser, IEEE J. of Quant. Electr. \textbf{38},
533 (2002).

\bibitem{2} A. Tredicucci, R. Kohler, L. Mahler, H.E. Beere, E.H. Linfield,
and D.A. Ritchie, Semicond. Sci. Technol. \textbf{20}, S222 (2005).

\bibitem{3} L.L. Bonilla and H.T. Grahn, Reports on Prog. in Phys. \textbf{68%
}, 577 (2005).

\bibitem{4} A. Wacker, Phys. Rep. \textbf{357}, 86 (2002).

\bibitem{5} S.A. Ktitorov, G.S. Simin, and V.Ya Sindalovskii, Sov. Phys.
Solid State, \textbf{13} 1872 (1971); A.A. Ignatov and Yu.A. Romanov, Phys.
Status Solidi B, \textbf{73} 327 (1976); A.A. Ignatov, E.P. Dodin, and V.I.
Shashkin, Mod. Phys. Lett. \textbf{B5}, 1087 (1991); A.A. Ignatov, K. F.
Renk, and E.P. Dodin, Phys. Rev. Lett. \textbf{70}, 1996 (1993).

\bibitem{6} N. V. Demarina and K. F. Renk, Phys. Rev. B 71, 035341 (2005);
V.N. Sokolov, L. Zhou, G.J. Iafrate, and J.B. Krieger, Phys. Rev. B 73,
205304 (2006).

\bibitem{7} Y. Shimada, K. Hirakawa, M. Odnoblioudov, and K. A. Chao, Phys.
Rev. Lett. \textbf{90} 046806 (2003); Y. Shimada, N. Sekine, and K.
Hirakawa, Appl. Phys. Lett. \textbf{84}, 4926 (2004);

\bibitem{8} N. Sekine and K. Hirakawa, Phys. Rev. Lett. \textbf{94} 057408
(2005); K. Hirakawa and N. Sekine, Physica E \textbf{32}, 320 (2006).

\bibitem{9} P.G. Savvidis, B. Kolasa, G. Lee, and S.J. Allen, Phys. Rev.
Lett. \textbf{92} 196802 (2004); P. Robrisha, J. Xub, S. Kobayashic, P.G.
Savvidis, B. Kolasa, G. Lee, D. Mars, and S.J. Allen, Physica E \textbf{32},
325 (2006).

\bibitem{10} H. Willenberg, G. H. Dohler, and J. Faist, Phys. Rev. B \textbf{%
67} 085315 (2003).

\bibitem{11} F.T. Vasko, Phys. Rev. B \textbf{69}, 205309 (2004).

\bibitem{12} F.T. Vasko and A.V. Kuznetsov, \textit{Electron States and
Optical Transitions in Semiconductor Heterostructures} (Springer, New York,
1998).

\bibitem{13} F.T. Vasko and O.E. Raichev, \textit{Quantum Kinetic Theory and
Applications} (Springer, New York, 2005).

\bibitem{13A} Such transformations of Eq. (7) give us 
\[
\frac{2}{L^{3}}\sum_{r\nu }f_{\varepsilon _{r\nu }} 
\]

\bibitem{14} Transformation from Eq.(8) to (9) is performed with the use: 
\[
\frac{f_{\varepsilon _{1}}-f_{\varepsilon _{2}}}{\mathcal{E}+\varepsilon
_{1}-\varepsilon _{2}}=\int d\varepsilon \int d\varepsilon ^{\prime }\frac{%
f_{\varepsilon }-f_{\varepsilon ^{\prime }}}{\mathcal{E}+\varepsilon
-\varepsilon ^{\prime }} 
\]%
\[
\times \delta (\varepsilon -\varepsilon _{1})\delta (\varepsilon ^{\prime
}-\varepsilon _{2}) 
\]

\bibitem{15} G.D. Mahan, \textit{Many-Particle Physics} (Plenum Press, N.Y.,
1990).

\bibitem{16} T. Ando, A.B. Fowler, F. Stern, Rev. Mod. Phys. \textbf{54},
437 (1982).

\bibitem{17} For the Gaussian case, $\gamma =0$, the convolution of spectral
densities takes the form: $D(\varepsilon ,\varepsilon ^{\prime})=\exp
\{-[(\varepsilon - \varepsilon^{\prime })/2\Gamma
]^{2}\}\{1-erf[(\varepsilon +\varepsilon ^{\prime })/\Gamma ]\}/4\sqrt{\pi }%
\Gamma $. Similar behavior, with a more complicate analytical function,
takes place for the Lorentzian case.

\bibitem{18} D.A. Dahl and L.J. Sham, Phys. Rev. B \textbf{16}, 651 (1977).

\bibitem{19} E. Scholl, \textit{Nonequilibrium Phase Transitions in
Semiconductors }(Springer, Berlin, 1987).

\bibitem{20} F.T. Vasko, P. Aceituno, and A. Hern\'{a}ndez-Cabrera, Phys.
Rev. B \textbf{66} 125303 (2002); F.T. Vasko, JETP \textbf{93} 1270, (2001).

\bibitem{21} S. Kohen, B.S. Williams, and Qing Hu, J. of Appl. Phys. \textbf{%
97}, 053106 (2005).

\bibitem{22} W. Zietkowski and M. Zaluzny, J. of Appl. Phys. \textbf{96},
6029 (2004).
\end{thebibliography}
\end{document}